\documentclass{llncs}
\usepackage{graphicx} % for pdf, bitmapped graphics files
\usepackage{amsmath} % assumes amsmath package installed
\usepackage{amssymb}  % assumes amsmath package installed
\usepackage{hyperref}
\usepackage{bigstrut}
\usepackage{tikz}
\usepackage{algorithm2e}
\usepackage{booktabs,caption}
\usepackage[flushleft]{threeparttable}
\usepackage{array}
\usepackage{multirow}
\begin{document}

\title{Its All in a Name: Detecting and Labeling Bots by Their Name }
\titlerunning{Random String Classification}  % abbreviated title (for running head)
\author{David M. Beskow \and Kathleen M. Carley}
\authorrunning{K. Carley et al.} % abbreviated author list (for running head)
\institute{School of Computer Science \\ 
Carnegie Mellon University \\ 5000 Forbes Ave, Pittsburgh, PA 15213, USA \\
Email: \email{dbeskow@andrew.cmu.edu} and  \email{kathleen.carley@cs.cmu.edu}}

\maketitle              % typeset the title of the contribution

\begin{abstract}
Automated social media \emph{bots} have existed almost as long as the social media environments they inhabit.  Their emergence has triggered numerous research efforts to develop increasingly sophisticated means to detect these accounts.  These efforts have resulted in a \emph{cat and mouse} cycle in which detection algorithms evolve trying to keep up with ever evolving \emph{bots}.  As part of this continued evolution, our research proposes a multi-model `tool-box' approach in order to conduct detection at various \emph{tiers} of data granularity. To support this toolbox approach this research also uses random string detection applied to user names to filter twitter streams for bot accounts and use this as labeled training data for follow on research.
\end{abstract}
\section{Introduction}

Automated social media accounts, often called ``bots'', are increasingly used on many social media sites.  Ever since social media sites built Application Programming Interfaces (API's) that allow their platforms to integrate with other platforms and applications, various actors have developed computer routines that conduct a variety of automated tasks on the respective social media ecosystems.  While some bots are designed for positive purposes \cite{graham2016socialbots}, many others range from nuisance (i.e. a spam bot) to propaganda \cite{lumezanu2012bias}, suppression of dissent \cite{verkamp2013five}, and network infiltration/manipulation \cite{freitas2015reverse,benigni2016tweets} .  They have recently gained wide-spread notoriety due to their use in several major international events, including the British Referendum known as ``Brexit'' \cite{howard2016bots}, the American 2016 Presidential Elections \cite{bessi2016social}, the aftermath of the 2017 Charlottesville protests \cite{AprilGlaser}, the German Presidential Elections \cite{neudert2017junk}, the conflict in Yemen \cite{yemen}, and recently in the Malaysian presidential elections \cite{ananthalakshmi_2018}.  

%Of particular interest is the use of bots to influence information and beliefs.  While commercialized spam bots arguably influence information flows (the general purpose of all marketing campaigns is to influence information and beliefs), the use of these same techniques by nation states, military, paramilitary organizations, and terrorist networks has far reaching ramifications.  It is for this reason that social media executives have been summoned to the legislative bodies of several countries to answer questions regarding automated behavior on their platforms \cite{congress}.

%Politicians and news organizations have recent criticized social media companies, pointing to the ease at which even junior researchers are able to develop bot detection programs \cite{undergrads}.  Others, however, are quick to point out that incorporating even the best of these detection algorithms into policy decisions becomes problematic \cite{thieltges2016devils}.  If the policy implication of bot detection means account suspension, then the accuracy of algorithms must be extremely high for these algorithms to be used in the implementation of this policy on platforms that could contain billions of accounts. On some of these platforms, a 1\% error rate affects over 10,000,000 accounts (note that at present, even the best teams have error rates of approximately 4-5\%, and these rates are generally with dated data).  

As these bots have proliferated and their use is being discussed broadly in the media and political bodies, researchers have increasingly developed methods to detect these accounts.  The same openness and ease of use of the social media API's that facilitates the creation and use of automated accounts also facilitates the collection of data used to detect them.  As detection efforts proliferate, bot engineers change and adapt in order to survive and succeed in a dynamic environment.  The requirement for higher accuracy in the midst of a changing \emph{signal} motivates our efforts to improve not only the models that detect bots, but the labeled data that is used to train them. 

This paper lays the foundation for a \emph{tiered} supervised machine learning approach to bot detection and characterization.  This approach acknowledges that there are discrete levels or `tiers' of data granularity, and seeks to develop supervised machine learning models for each \emph{tier} of data on social media platforms.  The resulting bot detection `toolbox' ensures that researchers have requisite models for their specific data granularity.  Some research aims at understanding bot behavior in large conversations (analyze overall bot presence in the Twitter conversation surrounding the 2018 mid-term elections), while other research aims at characterizing a handful of accounts (an depth analysis of the top 10 most influential followers of the NATO Twitter account).  A toolbox approach provides distinct models to support both requirements, allowing researchers to analyze the proverbial \emph{forest} with one model and \emph{trees} with a distinct but related model.

\begin{table}[htbp]
  \begin{threeparttable}
  \centering
  \caption{
  Four \emph{tiers} of Twitter data collection to support account classification}
    \begin{tabular}{|m{1cm}|m{3cm}|m{2cm}|m{3cm}|m{2.8cm}|}
    \hline
    \centering{Tier} & \centering{Description} & \centering{Focus} & Collect/process Time per 250  Accounts & \# of Data Entities (i.e. tweets) \\ \hline \hline
    Tier 0 & Tweet text or \newline{} user name & Semantics & N/A**  & 1 \\ \hline
       Tier 1 & \emph{User} Object \newline{}+ 1 Tweet object  & Account Meta-data  & $\sim 1.9$ seconds & 2 \\ \hline
      Tier 2 &  \emph{User} Object  \newline{}+ Timeline   & Temporal patterns & $\sim 3.7$ minutes & 200+ \\  \hline
     Tier 3 &   \emph{User} Object \newline{}+ Timeline \newline{}+ Friends Timeline  & Network patterns & $\sim 20$  hours & 50,000+ \\  \hline
    \end{tabular}%
  \label{tab:datasets}%
%   \begin{tablenotes}
%       \small
%       \item ** This tier of data collection was presented by \cite{kudugunta2018deep} and assumes the status text is acquired outside of the Twitter API  
%     \end{tablenotes}
  \end{threeparttable}
\end{table}%

To support the development of this toolbox, our research has identified several tiers of Twitter data collection and developed related machine learning feature space and models.  The constraints of data availability and rate limiting associated with the Twitter API \cite{rateLimiting} artificially create these tiers, which are summarized in Table \ref{tab:datasets}.  \emph{Tier 0} involves just a single entity, most often either a single \emph{status} or \emph{screen name}.  \emph{Tier 1} is the \emph{tweet} object (and associated \emph{user} object), and is the most common data granularity collected by researchers.  \emph{Tier 2} adds a user \emph{timeline} object (up to last 3,200 tweets) for every account, and \emph{Tier 3} adds the larger conversation that an account interacts with (i.e. an ego-network conversation).  

The data, feature space, and models associated with higher \emph{tiers} are rich and provide higher accuracy, but computationally expensive, as indicated in Table \ref{tab:datasets}. \emph{Tier 3} models can take over 20 hours to process 250 accounts.  Currently, researchers who use Tier 2 models on large datasets are forced to sample their data and assume that their sample is representative of overall bot distribution and characteristics. By providing models at Tier 0 and Tier 1, our toolbox will allow these researchers to conduct bot detection on 100\% of their data instead of sampling.

While numerous research efforts have attempted to exploit pieces and parts of this data spectrum, few have attempted to create a comprehensive approach that covers all tiers.  The closest effort that we've seen is the Botometer effort discussed later in this paper.  While offering an robust model through an accessible API, it is only offered at Tier 2, meaning high volume classification is computationally expensive.   Additionally, if does not exploit the rich network features available at Tier 3.  

This paper seeks to lay the groundwork for a toolbox approach to bot detection, discuss screen name focused data annotation, as well as build and evaluate a Tier 0 model.  See \cite{beskow2018botHunter} for discussion of our Tier 1 model and \cite{besow2018asonam} for initial efforts to develop baseline Tier2 and Tier 3 models.

Our work therefore makes three primary contributions to the literature.  First, we propose a novel random string detection model that is specifically designed to detect 15 character randomly generated strings.  When applied to the \emph{screen name} field of Twitter data, this technique is able to easily filter accounts that are likely bot accounts.  Second, by applying this filtering technique to a large sample collected from the Twitter Streaming API, we have produced a large and diverse annotated data set for use in training more robust specialized and general purpose bot detection models.  Finally, this paper lays the foundation for \emph{bot-hunter}, a multi-model `toolbox' approach to bot detection.

This paper begins with a brief description of the background of general bot detection, as well as past  efforts perform random string detection.  We will then describe the models and algorithms that we developed for random string classification, as well as methods that we used to evaluate them on the narrow tasks that they were created for.  Finally, we describe how we've applied this algorithm to create a large and diverse annotated Twitter bot data set for use by the research community.  

\section{Related Work}

\subsection{Twitter bot detection}
% Early work in bot development was ignited by Alan Turing's challenge to create a human-computer interaction in which the human can not distinguish whether or not they're interacting with a human or a computer \cite{turing1950computing}.  This challenge led to the creation of a chatbot called ELIZA by Joseph Weizenbaum in the 1960's \cite{weizenbaum1966eliza}.  This early effort was followed by numerous other chatbot efforts over the five decades.  Even with multiple research efforts in the area, it wasn't until the development and expansion of social media platforms that bot engineers had an ubiquitous environment where these bots could be deployed.  With the development and spread of social media, these bots have rapidly proliferated.

Although early work on classifying Twitter accounts dates back to as early as 2008 \cite{krishnamurthy2008few}, the deliberate detection of automated accounts on the Twitter Platform began in earnest in 2010 when \cite{chu2010tweeting} conducted three-class classification (human, bot, cyborg) using an ensemble model.  In 2011, a team from Texas A\&M became the first to use \emph{honey pots} to detect thousands of bots \cite{lee2011seven}.  These \emph{honey pots} used bots that generate nonsensical content, designed only to attract other bots.  The Texas A\&M bots attracted thousands of bots, and generated a labeled data set that has been used on many later research efforts.  This \emph{honey pot} method was repeated by others to create similar data sets in other parts of the world \cite{subrahmanian2016darpa}.

%This ensemble model was trained on hand coded data and leveraged temporal entropy measures, Bayesian classification (CRM114) for content base spam detection, and account properties.  Their decision maker algorithm was initially built on Linear Discriminant Analysis, but eventually used Random Forest in later work \cite{chu2012detecting}.  Other early works investigated automated accounts from the perspective of spam and spam prevention \cite{yardi2009detecting,grier2010spam,thomas2011suspended}.

%In 2013, \cite{wang2012social} explored the idea of Crowd Sourcing bot detection.  Although it showed some success, it was costly at scale, and generally required multiple workers to examine the same account.  This crowd-sourcing method has come under later criticism \cite{ferrara2016rise}.

In 2014, Indiana University and the University of Southern California launched the \emph{Bot or Not} online API service \cite{davis2016botornot}.  This used traditional classification models trained on the Texas A\&M dataset to help users evaluate whether or not an account is a bot.  \emph{Bot or Not} leverages \emph{network}, \emph{user}, \emph{friend}, \emph{temporal}, \emph{content}, and \emph{sentiment} features with Random Forest classification.    

In 2015 the Defense Advanced Research Projects Agency (DARPA) sponsored a Twitter bot detection competition that was titled ``The Twitter Bot Challenge'' \cite{subrahmanian2016darpa}.  This four week competition pitted four teams against each other as they sought to identify automated accounts that had infiltrated the informal Anti-Vaccine network on Twitter.  Most teams in the competition tried to use previously collected data (mostly collected and tagged with \emph{honey pots}) to train detection algorithms, and then leverage tweet semantics (sentiment, topic analysis, punctuation analysis, URL analysis), temporal features, profile features, and some network features to create a feature space for classification.  All teams used various techniques to identify initial bots, and then used traditional classification models (SVM and others) to find the rest of the bots in the data set.  

Most recently, the team from Indiana University re-branded \emph{Bot-or-Not} to \emph{Botometer}, increasing the set of features to 1,150 account related features \cite{ferrara2017measuring}.  Their team compared Random Forests, AdaBoost, Logistic Regression and Decision Tree classifiers and still found that Random Forests performed best.  They also attempted to update their training data by manually annotating tweet accounts, and merging this with the original Texas A\&M Dataset (collected in 2011).

The continued use of the 2011 Texas A\&M data highlights the difficulty that researchers have in creating and/or updating the labeled data that is used train algorithms to find these automated accounts.  The use of aging training data for bot classification also ensures that emerging bots are likely to avoid detection.  Additionally, since bots have a variety of purposes as well as a spectrum of actors that create/use them, the collection technique used for labeled data will bias the detection toward that family of bots.  For example, the \emph{honey pot} collection technique will bias toward bots that randomly follow accounts, but may not detect intimidation bots that conduct targeted following and messaging.

\subsection{Classifying algorithmic character strings}

Classifying strings as \emph{random} or \emph{not random} in order to filter or flag anomalous events has a limited background.  

Several methods have been proposed for identifying or highlighting the randomness of character strings.  Some have proposed leveraging Shannon's Entropy calculation \cite{shannon1948bell} as a method for sorting strings by a measure of randomness.  Some cyber security research teams have proposed a similar detection methods in order to detect domain names that are generated by Domain Generation Algorithms (DGA).  These teams have separately used Kullback-Leibler Divergence \cite{yadav2010detecting}, a dictionary approach \cite{defcon} and Markov modeling \cite{raghuram2014unsupervised}.

The past research most closely connected to our effort was conducted by LinkedIn in 2013.  At that time \cite{freeman2013using} presented the application of the Naive Bayes model on Character N-grams features of LinkedIn account names in order to identify \emph{spammy} accounts (first and last name as provided by the account owner).  This effort was very effective, and replaced the legacy spam detection models that LinkedIn was using on their OSN.  To date, our team has not found any team that has replicated a similar approach to Twitter screen names.

\subsection{Project background}

Our team has focused on detecting, characterizing, and modeling the behavior of bots, bot networks and their creators.  In doing this we've studied several recorded bot events. Recently we focused on a known and publicized bot attack against the Atlantic Council Digital Forensic Labs (DFR Lab), and tangentially against the NATO Public Affairs Office.  This attack primarily occurred between August 28 and August 30, 2017. We also focused on a recorded bot harassment event against journalists in Yemen \cite{yemen}.  In both events we observed numerous bot accounts that used 15 character randomly generated alpha-numeric strings for the screen name.  Examples of this include \textbf{Wy3wU4HegLlvHgC}, \textbf{5JSQavWW3tvQwA7}, and \textbf{gG6RKc6QBqOLKyU} (these are not real Twitter accounts).  Note that these randomly generated strings always sample from upper and lower case alpha-numeric characters.  Observing this phenomenon motivated the construction of this algorithm and its application on Twitter at large in order to observe other bots and bot actors that are using these same type of bot screen names. More importantly, we hope this dataset can be used as a large and diverse annotated bot training data for larger and more comprehensive machine learning models.

\section{Modeling}
\subsection{Feature engineering}

In order to develop a random string detection model for this unique case, we constructed training data consisting of 200,000 non-random Twitter screen names (randomly sampled from Twitter and manually verified as non-random) and 200,000 randomly generated 15 digit strings.  We then developed a combination of heuristic filtering and traditional machine learning models to label the string as \emph{random} or \emph{not random}.  This development is described below.

For feature engineering, the primary feature that we extracted from the strings was character n-gram.  For string $s$ with length $m$, a character n-gram is the $(m-n+1)$ sequential substrings of length $n$ found in string $s$.  In our case, we explored several settings for $n$, to include using multiple values in the same feature set (i.e. using both bigrams and trigrams). 

We then transformed the resulting sparse character n-gram matrix using term frequency-inverse document frequency (TF-IDF).    TF-IDF is defined in Equation 1 and 2 below, and is used to scale the characters by the information that they provide.  In our case, frequent characters in a string provide information, but not if they're frequent in all of the strings.  To calculate the IDF for character $c$ in strings $s$, we take the logarithm of the ratio of  the total number of strings in corpus $S$ by the number strings that contain $c$, as shown in Equation \ref{eq:1}.

\begin{equation}\label{eq:1}
	idf(c,S)=log \frac{N}{\lvert \lbrace   c \in S : c \in s \rbrace \rvert } 
\end{equation}

We then calculate the TF-IDF for character $c$ in string $s$ found in corpus $S$ as follows

\begin{equation}
tfidf(c,s,S) = tf(c,s) \dot idf(c,S)
\end{equation}

This therefore weights characters that have a high local frequency but a lower global frequency.  At first it may seem that TF-IDF is unnecessary since each character n-gram is equally likely in random strings, given a strong pseudo-random number generator.  n-grams are not equally likely for human generated strings, however.  Given this fact we felt it appropriate to transform the data with TF-IDF.

These features were merged with several other features. We started by merging the normalized count of upper case, lower case, and numeric characters.  n-gram generation by default converts all text to lower case. We maintained this default behavior, but saw that the number of upper and lower case in letters in particular provided a strong signal.  Since our training data contained some human generated strings that were not 15 characters in length, we normalized these counts.  

Additionally, we included the Shannon string entropy in our feature set. Shannon string entropy, while not strong enough to use by itself in our case, still provides a strong signal that we felt would be useful. We will test this assumption below. Shannon entropy is defined in \ref{eq:3}, where $p_i$ is the normalized count for each character found in the string.

\begin{equation}\label{eq:3}
H \left(A \right) = - \sum_{i=1}^{n}p_i log_2 p_i
\end{equation}

The A full table of features is given in Table \ref{feature-table}.

\begin{table*}[h]
\centering
\caption{Features for Random String Detection}
\label{feature-table}
\begin{tabular}{p{3.5cm}p{2cm}p{6.5cm}}
\hline 
Feature & Type&  Description  \bigstrut \\ \hline \hline 
Character Bi-gram & Numeric & Term frequency inverse document frequency of bi-gram \bigstrut \\ 
No. lower case & Numeric  &  Normalized count of lower case letters   \\ 
No. upper case & Numeric   &  Normalized count of upper case letters    \\
String entropy  & Numeric  & Shannon String entropy   \\
\hline
\end{tabular}
\end{table*}

We used the $scikit-learn$ package \cite{scikit-learn} to explore and build the machine learning classification model for Random Strings.  We evaluated Naive Bayes, Logistic Regression, and Support Vector Machines (SVM) with 10 fold cross-validation. The results are presented in Table \ref{string-pred}.  We conducted model comparisons between these models, and found SVM and Logistic Regression did are not statistically different ($t = 0.62912$, $df = 18$, $p.value < 0.5372$). Given these results, we used Logistic Regression for our production model, given that it is simpler and faster.  Note that this result entails significantly more training data than we used in earlier research (see \cite{beskow2018random}), where SVM performed better.

\setlength{\tabcolsep}{6pt}
\begin{table}[htbp]
  \centering
  \caption{Model Performance in Classifying Randomly Generated Strings for Screen-names}
    \begin{tabular}{lcccccc}
    \hline
    Model & Accuracy & F1 & Kappa & Precision & Recall & ROC  AUC \bigstrut \\
    \hline
    \hline
    Log. Regression & 0.996 & 0.996 & 0.991 & 0.994 & 0.997 & 0.999 \bigstrut\\
    \hline
    Na\"ive Bayes & 0.969 & 0.97  & 0.939 & 0.947 & 0.995 & 0.996 \bigstrut\\
    \hline
    SVM   & 0.996 & 0.996 & 0.993 & 0.995 & 0.998 & 1 \bigstrut\\
    \hline
    \end{tabular}%
 \label{string-pred}
\end{table}%

Before predicting whether or not a string was random, we first applied several heuristic filters.  These verified that 1) the string was 15 characters in length, and 2) contained at least one capital letter, lower case letter, and numeric digit.  This final filter was applied given that 15 character strings have a 0.02\% chance of not containing a capital or lower case letter and a 7\% chance of not containing a numeric digit. This heuristic was applied given that precision was a higher priority than recall.   

% The final algorithm is provided in Algorithm \ref{my_alg}.

% \begin{algorithm}[h]
% \SetAlgoLined
%  \If{length equals 15 characters}{
%  upper = number of upper case letters\;
%  lower = number of lower case letters\;
%   \If{$upper \times lower \neq 0$}{
%    predict randomness with SVM\;
%    \If{random}{
%    	write to file
%    }
%    }{

%   }
%  }
%  \caption{SVM model wrapped in heuristic filters}
%  \label{my_alg}
% \end{algorithm}

\begin{figure}[h]
  \includegraphics[width=\columnwidth]{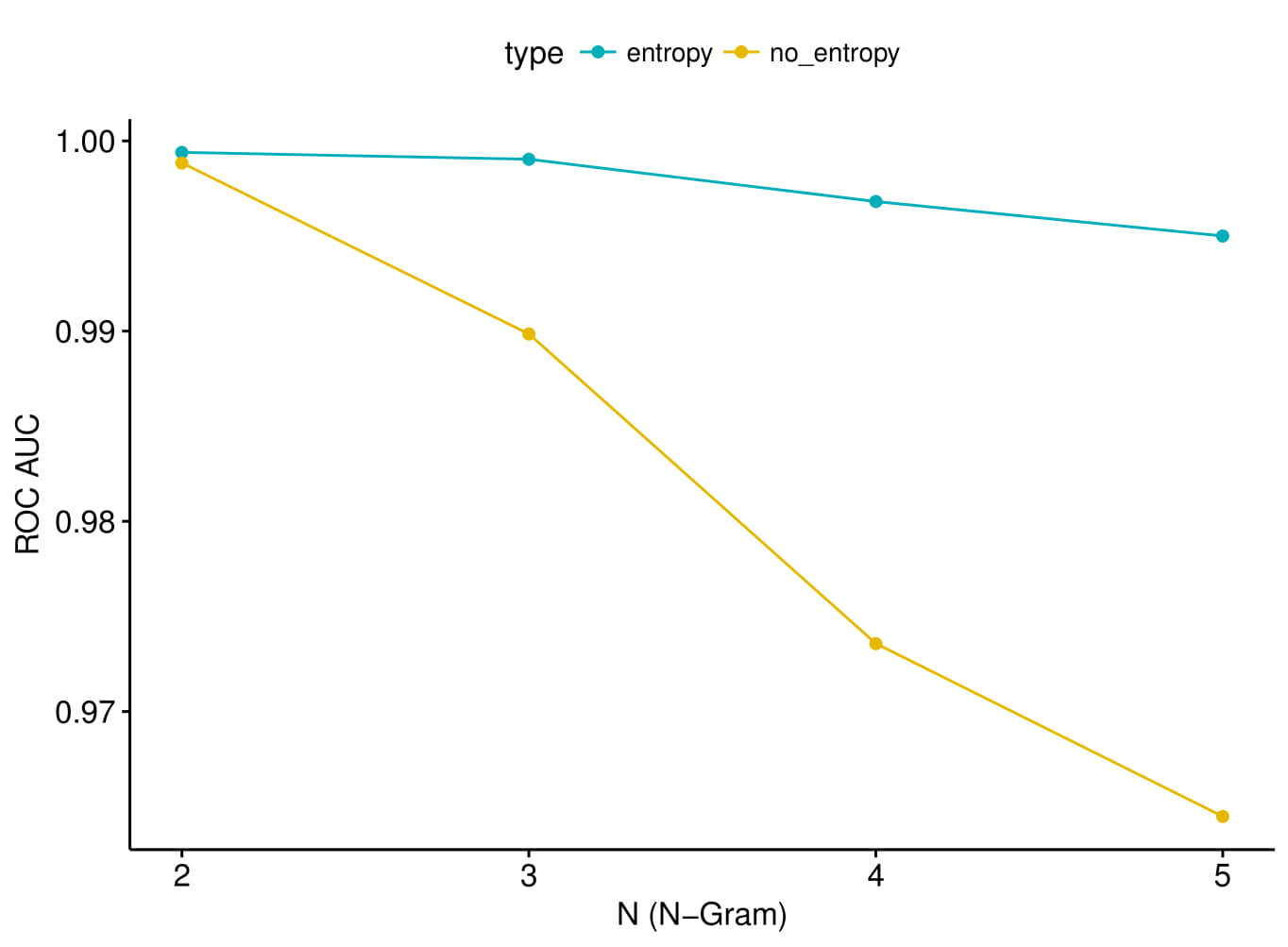}
  \caption{Evaluating n (number of characters in n-gram) and use of Shannon's entropy as a feature}
  \label{learning}
\end{figure}

In Figure \ref{learning} we evaluate the best value of $n$ (number of characters for n-gram) as well as whether or not using Shannon's Entropy as a column feature provides leverage in prediction.  In this visualization we see that bigrams with Shannon's entropy provides the best leverage in predicting random strings.  

In addition to exploring the feature based machine learning models discussed above, we also explored the use of Markov model of character sequencing, but found during initial exploration that this did not have sufficient power to classify the strings given the inherent random nature of human generated screen names. Additionally, we explored using Shannon entropy as the only measure for filtering these strings.  Once again, while helpful, this method did not demonstrate sufficient power for our purposes.

\subsection{Model Deployment}

Our primary use for the algorithm was to filter accounts with 15 character random strings from a Twitter data stream.  To do this we ran a random sample from the Twitter Streaming API from 23 December 2017 to 20 June 2018.  During this time the stream collected approximately 433 million tweets.  This collection was done without any semantic or geographic filters, and stored the raw JSON files that are returned by the Twitter API.

Having performed the collection, we next applied our algorithm to all 433 million tweets, filtering out all accounts that were labeled as having 15 digit randomly generated screen name.  This produced a collection of 7.8 million tweets from 1.7 million unique accounts.

\section{Model Evaluation}

Given the desired use case of annotating diverse bot accounts, we conducted two evaluations on our results.  First, we wanted to estimate the false positive rate on our random string detection, since false positives have a high likelihood of not being an autonomous account.  To accomplish this we randomly selected 1,000 of the screen names that were labeled as random, and manually identified those that contained clear words or acronyms.  Given this method, we estimate that our false positive rate is approximately 1\%.

Additionally, we wanted to estimate the percentage of random character screen name accounts that are automated, or appear automated.  In other words, how many of our true positive random string accounts are truly bots. To estimate this, we randomly sampled 100 accounts, verified that the user name appeared random, and inspected the account in the Twitter web client.  Of the 100 that we manually inspected, five were suspended, eight provided no results (most likely the account was closed by the user), and all others exhibited autonomous behavior.  After thoroughly evaluating these 100 randomly sampled accounts we were were satisfied that this methodology provides annotated bot data that is at least as accurate as \emph{honey pot} data, and likely has a wider range of bot types.  

\subsection{Data Characterization}

One of our first tasks in exploring the data is to understand how these accounts differ from the average Twitter account, and whether those differences were uniform across the language of the bot creator.  

% In general, autonomous accounts produce tweets at a much higher volume and rate than human actors.  The mean number of tweets for our accounts was 7,918 (median = 1,125).  In general bots have a low number of followers (most people don't follow bots), but they tend to follow many accounts, trying to build influence.  Following this pattern, the median number of followers is in our data set is 55, but the median number of accounts they follow is 130.   

99\% of the 7.8 million tweets in this dataset are associated with seven languages. It's interesting to note that none of the Continental European Languages (French, Spanish, German, Portuguese, Italian, etc) are in this list.   Somewhat surprisingly, the proportion associated with Japanese and Arabic accounts is very high, second only to English.  A full breakdown of the languages and a short general description of our observations are provided in Table \ref{tab:results}.  Only 840 tweets contained coordinate locations, and these locations are strongly correlated to the languages mentioned below (United States, Japan, the broader Middle East, Russia, and Thailand).

The major observations from Table \ref{tab:results} are that the random string accounts are younger, less popular, and less active than the average Twitter account.  We see that the median age for the random bots is 224 days, compared to 1,248 days for your average active Twitter account.  The median number of followers/friends ratio for the random string bots is 6/39 versus 277/294 for the average Twitter account.  We also see that the median random string bot account only produced 54 tweets over its lifetime, versus 8,216 for the average account (this comparison is affected by age difference).

% \begin{table}[h]
% \centering
% \caption{Characterization by Language}
% \label{lang}
% \begin{tabular}{p{2cm}p{1.5cm}p{7.5cm}}
% \hline
% Language & \% Total  & General description  \bigstrut \\ \hline
% Japanese & 184,385 & High concentration of anime media sharing \bigstrut \\
% Arabic & 111,523 & High percentage of young accounts, some automated Koran passage sharing \\
% English & 94,804 & Contains a high number of non English hash tags  \\
% Korean & 41,870 & Varied \\
% Thai & 14,195 & High concentration of adult content   \\
% Russian & 13,461 & Varied  \\
% \hline

% \end{tabular}
% \end{table}

\setlength{\tabcolsep}{1pt}
% Table generated by Excel2LaTeX from sheet 'random_descriptive_stats_table'
\begin{table}[h]
  \begin{threeparttable}
  \centering
  \caption{Summary Statistics by Language}
    \begin{tabular}{cccccccccc||c}
    \hline
    \multicolumn{2}{c}{Language} & \multicolumn{1}{l}{Arabic} & \multicolumn{1}{l}{English} & \multicolumn{1}{l}{Japanese} & \multicolumn{1}{l}{Korean} & \multicolumn{1}{l}{Russian} & \multicolumn{1}{l}{Thai} & Chinese & other & Normal*  \\
    \multicolumn{2}{c}{\# of Accounts} & 246K & 626K & 593K & 103K & 61K & 47K & 21K & 18K &1599K\\ \hline
    \multicolumn{1}{c}{\multirow{7}[0]{*}{Age}} & min   & 61    & 61    & 61    & 61    & 62    & 61    & 62    & 61 & 6 \\
          & 25\% & 181   & 105   & 214   & 193   & 162   & 167   & 186   & 192 &  487\\
          & 50\% & 264   & 165   & 361   & 260   & 292   & 246   & 288   & 297 &  1,248\\
          & 75\% & 413   & 213   & 570   & 427   & 365   & 383   & 423   & 626 &  2,235\\
          & max   & 3,046 & 17,763 & 3,731 & 3,020 & 3,075 & 3,306 & 3,431 & 3,662 &  4,421\\
          & mean  & 326   & 210   & 449   & 342   & 322   & 310   & 357   & 550 &  1,412\\
          & std   & 216   & 253   & 315   & 229   & 228   & 219   & 247   & 609 & 1,008\\  \hline
    \multicolumn{1}{c}{\multirow{7}[0]{1.5cm}{\centering Followers Count}} & min   & 0     & 0    & 0     & 0     & 0     & 0     & 0     & 0 & 0 \\
          & 25\% & 3     & 0     & 2     & 0     & 0     & 1     & 0     & 2 &  78\\
          & 50\% & 15    & 2     & 19    & 2     & 1     & 5     & 1     & 15 &  277\\
          & 75\% & 63    & 17    & 108   & 10    & 5     & 24    & 6     & 85 &  818\\
          & max   & 828K & 1087K & 1322K & 23,681 & 54K & 177K & 50K & 944K & 40,550K \\
          & mean  & 171   & 61    & 136   & 32    & 93    & 163   & 97    & 295 & 3376 \\
          & std   & 2,716 & 2,054 & 2,366 & 268   & 1,013 & 2,044 & 921   & 7,581 &  94990 \\ \hline
    \multicolumn{1}{c}{\multirow{7}[0]{1.5cm}{\centering Friends Count}} & min   & 0     & 0    & 0     & 0     & 0     & 0     & 0     & 0 &  0\\
    		& 25\%& 26    & 5     & 10    & 1     & 6     & 31    & 7     & 29 &  118\\
          & 50\% & 79    & 26    & 49    & 21    & 31    & 88    & 32    & 91  & 294\\
          & 75\% & 226   & 73    & 168   & 74    & 53    & 258   & 70    & 242  & 695\\
          & max   & 640K & 349K & 75K & 25K & 17K & 18K & 12K & 88K  & 2,441K\\
          & mean  & 297   & 101   & 178   & 92    & 130   & 257   & 98    & 298  & 1044\\
          
          & std   & 1,543 & 682   & 482   & 326   & 594   & 498   & 322   & 1,034  & 8227\\  \hline
    \multicolumn{1}{c}{\multirow{7}[0]{1.5cm}{\centering Tweet count}} & min   & 1     & 1    & 1     & 1     & 1     & 1     & 1     & 1  & 1\\
          & 25\% & 15    & 6     & 25    & 26    & 16    & 24    & 15    & 19  & 1813\\
          & 50\% & 71    & 20    & 117   & 134   & 69    & 99    & 59    & 109  & 8216\\
          & 75\% & 319   & 83    & 515   & 601   & 234   & 370   & 286   & 627  & 27318\\
          & max   & 532K & 806K & 994K & 228K & 114K & 570K & 106K & 304K  & 16,176K\\
          & mean  & 819   & 301   & 930   & 934   & 456   & 684   & 517   & 1,727 & 26,652 \\
          & std   & 3,753 & 3,180 & 4,301 & 3,652 & 2,226 & 4,195 & 2,036 & 7,981  & 66,180\\ \hline
    \end{tabular}%
      \begin{tablenotes}
      \small
      \item * \emph{Normal} Twitter Accounts were sampled from the Twitter Streaming API
    \end{tablenotes}
  \label{tab:results}%
    \end{threeparttable}
\end{table}%

% The mean age of the accounts is 329 days, with 50\% of the accounts created in the last 224 days and 72\% of the accounts younger than 1 year old.  The relative young age of these accounts is highly indicative of their automated behavior.  The oldest accounts are associated with English account settings, and date back to summer 2007.

While some languages (Arabic, Japanese, Korean, and Thai) appear to be slightly more popular and active, in general these random string accounts appear to have a high number of accounts that are dormant, or at least in a state of low activity.  Some of these may be waiting to be activated for a given event or task, while others may be used for intimidation attacks (as some of these were with the Yemen journalist discussed above).  Intimidation accounts (accounts that follow a user in mass) do not need to be active or popular.  Their intent is to push another account out of the Twitter conversation through intimidation.

Given the fact that our data set contains primarily bot accounts, we observed a number of account suspensions during the course of our study.  Between mid December 2017 and August 22 2018, 247,022 accounts ({\raise.17ex\hbox{$\scriptstyle\mathtt{\sim}$}}15\%) were suspended by Twitter, while 46,985 accounts ({\raise.17ex\hbox{$\scriptstyle\mathtt{\sim}$}}2.7\%) were removed by the user. As the media and politicians put pressure on Social Media companies, the natural response is to increase their policing of this automated behavior on their platforms.  

\section{Conclusion}

% Our method effectively provides one of the largest and most diverse bot data sets created to date.    This labeled data is now available to conduct feature engineering and model evaluation, developing bot detection algorithms that can be generalized or specialized and can detect bots beyond those with random screen names.  

% This work also provides a rich data set that can be used for focused research on specific bot behaviors and their effect on networks and beliefs.  Our team has already began enriching the data by collecting the timelines and network relationship of subsets of the data.  

Research in this area is limited by a rich enough data set that supports identification of the wide range of types of bots, and that is sufficient to support studies of bot-evolution.  While the data used herein begins to address this issue, it is by no means comprehensive and needs further expansion.  We are working on such expansion.  However, restrictions on data sharing make it difficult to share this data.  Consequently, we are also working on data format that can be shared.  

Bots are part of the conversation in social media.  But not all bots are the same.  They vary in what they do, how they do it, and intent.  While some bots act independently others work in concert and still others are part of a cyborg - a human-bot partnership.  Research is needed to characterize types of bots and their evolution.  Research is also needed to identify the mapping between types of bots in use and types of information maneuver or social-group creation that, that type of bot supports or thwarts.

\section{Future Work}

Our future effort begins with the exploration of this dataset so that we can cluster these accounts by type and function.  We then intend to develop and train several specialized as well as a general purpose bot detection algorithms for use in detecting and classifying bots.  Once complete, our effort will shift to the detection and characterization of bot networks and the actors behind them.


%%%%%%%%%%%%%%%%%%%%%%%%%%%%%%%%%%%%%%%%%%%%%%%%%%%%%%%%%%%%%%%%%%%%%%%%%%%%%%%%

%%%%%%%%%%%%%%%%%%%%%%%%%%%%%%%%%%%%%%%%%%%%%%%%%%%%%%%%%%%%%%%%%%%%%%%%%%%%%%%%

% \appendices

% \subsection{Appendix 1: Convolutional Neural Network Learning}

% The plot of the CNN learning history for the direct classification of profile images (as \emph{bot} or \emph{human}) is shown in Figure \ref{learning}.  As seen here, the CNN struggles to find distinguishing characteristics, producing a shallow (and sometimes negative) learning rate. 

\section*{ACKNOWLEDGMENT}

This work was supported in part by the Office of Naval Research (ONR) Multidisciplinary University Research Initiative Award  N000140811186 and Award N000141812108,  the Army Research Laboratory Award W911NF1610049, Defense Threat Reductions Agency Award HDTRA11010102, and the Center for Computational Analysis of Social and Organization Systems (CASOS). The views and conclusions contained in this document are those of the authors and should not be interpreted as representing the official policies, either expressed or implied, of the ONR, ARL, DTRA, or the U.S. government.

%%%%%%%%%%%%%%%%%%%%%%%%%%%%%%%%%%%%%%%%%%%%%%%%%%%%%%%%%%%%%%%%%%%%%%%%%%%%%%%%

\bibliographystyle{plain}
\bibliography{botBibliography.bib}

\end{document}